# A Facile Approach to Prepare Self-Assembled, Nacre-Inspired Clay/Polymer Nano-Composites

Peicheng Xu, Talha Erdem and Erika Eiser*

*Cavendish Laboratory, University of Cambridge, Cambridge CB3 0HE, UK*

## Abstract

Nature provides many paradigms for the design and fabrication of artificial composite materials. Inspired by the relationship between the well-ordered architecture and biopolymers found in natural nacre, we present a facile strategy to construct large-scale organic/inorganic nacre-mimetics with hierarchical structure via a water-evaporation self-assembly process. Through hydrogen bonding, we connect Laponite-nanoclay platelets with each other using naturally abundant cellulose creating thin, flexible films with a local *brick-and-mortar* architecture. While the aqueous solution displays liquid crystalline textures, the dried films show a pronounced Young's modulus (9.09 GPa) with a maximum strength of 298.02 MPa and toughness of 16.63 MJm$^{-3}$. In terms of functionalities, we report excellent glass-like transparency along with exceptional shape-persistent flame shielding. We also demonstrate that through metal ion-coordination we can further strengthen the interactions between the polymers and the nanoclays. These ion-treated hybrid films exhibit further enhanced mechanical, and thermal properties as well as resistance against swelling and dissolution in aqueous environments. We believe that our simple pathway to fabricate such versatile polymer/clay nanocomposites can open avenues for inexpensive production of environmentally friendly, biomimetic materials in aerospace, wearable electrical devices, artificial muscle, and food packaging industry.

# Introduction

In the 21st century, there has been a major scientific challenge and increasing demand in reducing resource consumption and pollution emissions. This is extraordinarily prominent in shifting our reliance from traditional fossil fuel derived polymers to environmentally friendly materials made from bio-based polymers.[1,2] Nevertheless, due to its low stiffness and strength, simply using biopolymers is not enough to produce high-performance materials with multi-functionalities. Thankfully, natural load-bearing composites provide inspirational examples for innovative engineering designs, found in materials such as mother of pearl (nacre), spider silk, wood, teeth, insect cuticles and bone. These materials combine synergistically high strength, stiffness and toughness, which is provided by the well-ordered organization of the underlying micro and nanostructure.[3–5]

Among all natural composites, nacre has received intensive attention for its outstanding mechanical performance.[6,7] Like ceramics, nacre is strong and tough, arising from the slow growth of its *brick-and-mortar* architecture, which is composed of 2D inorganic hexagonal micro-platelets (95%) that are glued to each other by a small amount of organic fibrillar polymer matrix.[8–11] Due to the self-assembly of hard reinforcing building blocks and soft energy adsorbing layers as well as the sophisticated interfacial dynamics between them, its toughness is also 1000 times higher than its individual components.[12] Inspired by this, many research groups attempt to mimic nacre by filling the natural polymer matrix with nanoscale additives for the purpose of duplicating its exceptional properties.

So far, several systematic studies have been conducted in order to reproduce the complexity of nacre. Different types of nacre-inspired nanocomposites have been fabricated with inorganic brick nanosheets as the hard phase, such as molybdenum disulfide ($MoS_2$),[13] zinc oxide,[14,15] alumina oxide,[16] graphene oxide (GO),[17,18] layered double hydroxides (LDH),[19] montmorillonite (MMT)[20,21] and flattened double-walled carbon nanotubes (CNT).[22,23] Whereas, poly(vinyl alcohol) (PVA),[21] chitosan,[24] poly(L-lysine)-g-poly(ethylene glycol) (PLL-PEG),[25] poly-(methyl methacrylate) (PMMA),[15,26] and poly(acrylamide) (PAM)[18] were chosen as the flexible soft phase, acting as mortar between the inorganic building blocks. At the same time, several manufacturing strategies have been employed to prepare nacre-inspired materials. Using sequential deposition processes such as spinning-coating or layer-by-layer (LBL) deposition of polymers and nanoparticles can be used to achieve layered structure with controlled size but these techniques are time-consuming, inherently costly to prepare and difficult to scale up.[27] Employing a filtration strategy, though it is easy to operate, also suffers from scaling-up challenges for industrial processing because of the limited size of filtration setups.[17] Another approach is freeze casting, which is well known for being environmentally friendly but suffers from laborious multistep procedures and extensive consumption of energy.[28] Compared with LBL, filtration and freeze casting, electrophoretic deposition demonstrates high controllability but like the other methods, there still remains a major challenge for large-scale production.[29] In addition, in order to satisfy the ever-growing demand in the application world, bio-inspired materials with multi-functions should be considered. Here we present a new type of composite with a *brick-and-mortar* structure similar to that in nacre, but with very different characteristic lengths. We show that we can produce thin, flexible films with high transparency, outstanding mechanical properties, and tunable water-resistance diminishing fouling and fire-retardance. These films can be produced cost- and energy- effectively on large scale, using clays and cellulose which are abundant in nature, presenting a new alternative to plastics.

The clay we use here is Laponite, a smectite clay that is extensively utilized for structural materials in diverse scientific and industrial fields due to its stiffness and low cost.[29] Despite its high aspect ratio and limited solubility in water it has been challenging to access the lyotropic liquid crystalline phases of these systems, since the surface functional groups like hydroxyl and silanol groups render the flat surfaces negatively charged and the rims positively charged leading to strong aging and gelation. Therefore Laponite is an ideal additive for making nanocomopiste materials and drilling fluids. However, clays alone are brittle even under low loading levels; a toughening mechanism is therefore needed.[30]

Cellulose, a natural polysaccharide, is well known as most abundant organic macro-molecule on earth. Unlike petroleum-based polymers, cellulose is cheap, renewable, nontoxic, easily formable and often a side product in many industrial processes. Therefore, as a sustainable alternative, it has received extensive attention in environmental protection.[31] Sodium carboxymethyl cellulose (CMC) is a cellulose derivative with a few hydroxyl groups substituted by carboxymethyl groups.[32] Owing to the presence of substantial hydroxyl and carboxyl groups, it is easy for CMC to form hydrogen bonding with other constituents, serving as a load-adsorbing phase. And like alginate, CMC contains carboxylate ions, which means it can also be cross-linked by chelation with divalent cations.[33–35]

For the purpose of further strengthening the polymer-clay interface and therefore optimizing the material properties, metal ions are used for coordination.[36] Compared with other metal ions, calcium ion has attracted great interest due to its low cost, easy obtainability, low toxicity, relatively high solubility in water, good chemical durability and prominent fabricabilit.[33] In addition, using calcium ions means we can mimic nacre in a better way because calcium element happens to be the main component of natural nacre, which is composed of a large amount of aragonite (calcium carbonate).[9] Nevertheless, it is a pity that nacre-mimetic materials fabricated with Laponite and calcium-ion-coordinated CMC have rarely been reported.

In this paper, inspired by the structure of nacre, we report a facile and environmentally friendly way to fabricate the nanocomposite materials by using Laponite and CMC hybrid building blocks. The Liquid crystalline structure can be achieved when clay suspension is mixed with polymer solution without stirring. When finishing evaporating of the hybrid suspension, polymer-coated clay particles self-assembly into a highly oriented architecture. The resultant artificial material demonstrates high transparency and flexibility. By varying the weight ratio between CMC and Laponite, we obtain good mechanical properties and derive the fracture mechanism corresponding to the amount of clay contents. For the purpose of further enhancing the interfacial interactions and expanding the applications, calcium ions, as a crosslinking agent, are infiltrated into hybrid films. After introducing calcium ions, the Young's modulus, maximum strength and toughness can reach up to 9.09GPa, 298.02 MPa and 16.63 MJm$^{-3}$ respectively. It is also worth noting that we manage to achieve a balance between the high toughness and high stiffness. Apart from outstanding mechanical properties, our nanocomposite films also exhibit outstanding water-resistance and thermal stability as well as remarkable fire-retardant property.

## Results and Discussion

### Liquid Crystalline Gel

Aqueous Laponite solutions are colloidal dispersions, consisting of electriferous disk-like clay particles with an approximate diameter of 25nm while the thickness is fixed to 1nm by the covalently bonded 2D-crystal structure of this smectic silicate.[37,38] Due to the electrostatic attraction between rims and surfaces,[37] Laponite suspensions have been found to evolve from an initial liquid-like state to an arrested non-ergodic solid, with both liquid and gel phases showing continual aging (Figure. 1a). The transition of purely aqueous solutions of Laponite into either a transparent gel or glass only depends on the clay concentration and the ionic strength of the system.[39,40] For the purpose of avoiding a direct interaction between the clay particles that is leading to this gelation, we added carboxymethyl cellulose as possible polymeric stabilizer that could suppress the aging process. In Figure 1 we show a schematic drawing of both the Laponite particles and the CMC. When mixing the clay particles with the CMC chains, hydrogen bonds can form between hydroxyl groups (-OH) on the CMC chains and silanol (Si-O) groups on the negatively charged Laponite surfaces. Here we used carboxymethyl cellulose with a molecular weight of 700 kDa and estimated radius of gyration $R_g$ of around 100 nm,[41] which means the diameter of our CMC coils is 8 times larger than the average Laponite diameter. This means a single CMC chain can not only form several contact points with a single Laponite particle but bridge several particles or wrap completely around one particle, depending on the particle to polymer ratio and the total concentration. Therefore, when the polymer concentration is high enough, the polymer will bridge several Laponite particles and thus form an organic-inorganic highly viscous network. The desired coating method is achieved by first preparing separate clay and polymer solutions, which are then added to each other. We find that the sequential order of adding clay and polymer matters, as shown in Figure 2.

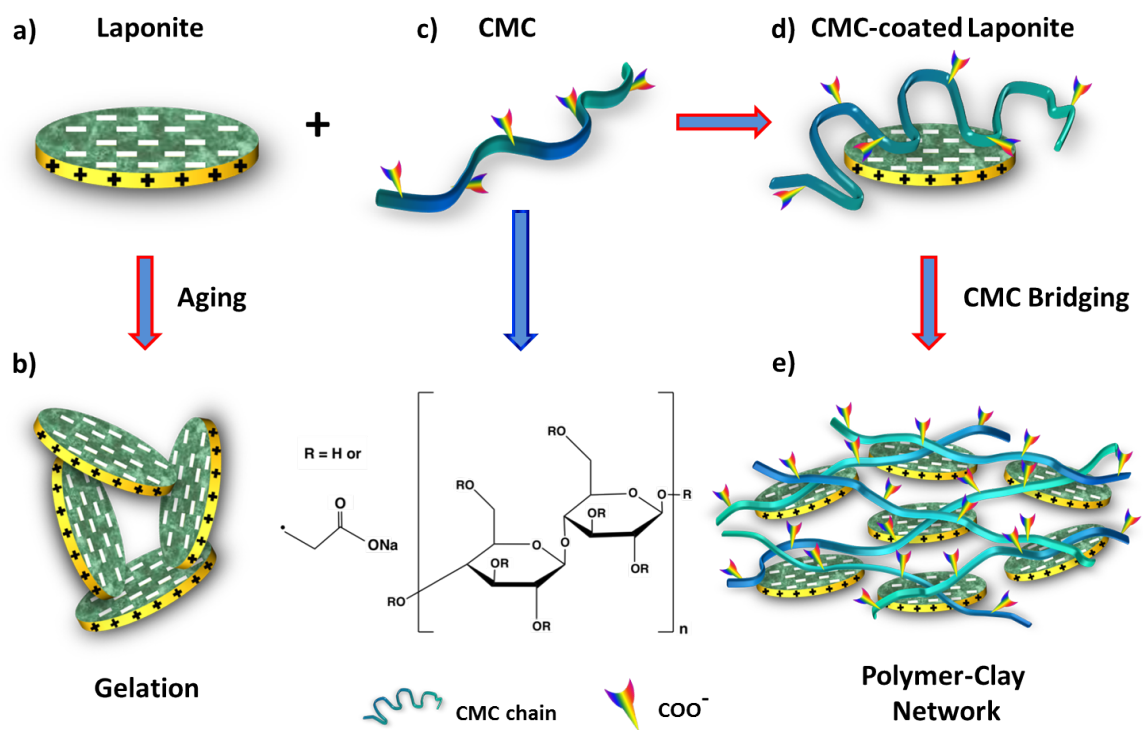

**Figure 1: (a)** A single Laponite disk with positive charges on the rims and negative charges on the surfaces. **(b)** Gelation due to electrostatic attraction between rims and surfaces.[37,40] **(c)** Chemical structure of CMC. **(d)** Schematic representation of CMC polymer coating onto individual Laponite surfaces through hydrogen bonding. **(e)** A homogeneous polymer-clay network in water formed through CMC bridging under the condition of extra CMC.

When we carefully placed the 0.8 wt% CMC solution on top of a 2.0 wt% Laponite solution, ensuring that there would be little mixing of the resulting liquid-liquid interface, a gel with colorful birefringence at the bottom of the closed vial was observed after a day, when viewed between two crossed polarizers (Figure 2a). Both clay and polymer dispersions were initially transparent and non-birefringent, and the interface between them remained relatively sharp indicating that indeed no mixing took place. However, the location of the interface started shifting considerably over the course of one day, reducing the lower pure Laponite volume as shown in Figure 2. When the order was reversed such that the more dense Laponite (density of pure Laponite is 2.53 g cm$^{-3}$)[42] solution was placed on top of the viscous but less dense CMC (density of pure CMC is 1.60 g cm$^{-3}$)[43] solution an asymmetric birefringent texture appeared, indicating that the Laponite solution slowly crept along the container walls underneath the CMC solution, thereby disturbing the liquid-liquid interface slightly but not destroying it. And again a birefringent texture of the Laponite rich layer appeared after a day. We argue that this birefringence is caused by the interplay of two effects: The much larger CMC coils can quickly form point-like adhesions with the clay particles close to the interface (Figure 1e), forming a highly viscous interfacial layer that prevented the two solutions to mix purely due to diffusion. Moreover, due to sodium and carboxylate ions, CMC dispersed in water behaves like a polyelectrolyte. Consequently an osmotic pressure gradient build up between the two liquid phases leading to the transport of water molecules into the CMC rich phase. As a result, the volume fraction of Laponite increased, causing a gradual orientational order of the confined clay particles in a colorful birefringence gel while the CMC region was still liquid. This was verified by tilting the sample tube (not shown here). As mentioned above the same phenomenon was observed when reversing the order with the difference that the denser Laponite solution was gliding underneath the initially lower CMC sample. Also note that CMC rich solutions did not show any birefringence at all concentrations used, while the birefringent Laponite gel can be analyzed in terms of a nematic phase using the Michel-Levy interference color chart.[44] Taking a Transmission Electron Microscope (TEM) image from the resulting liquid supernatant prepared with the CMC solution on top showed compete absence of Laponite particles, while some clay particles were entering the CMC phase when using the second approach.

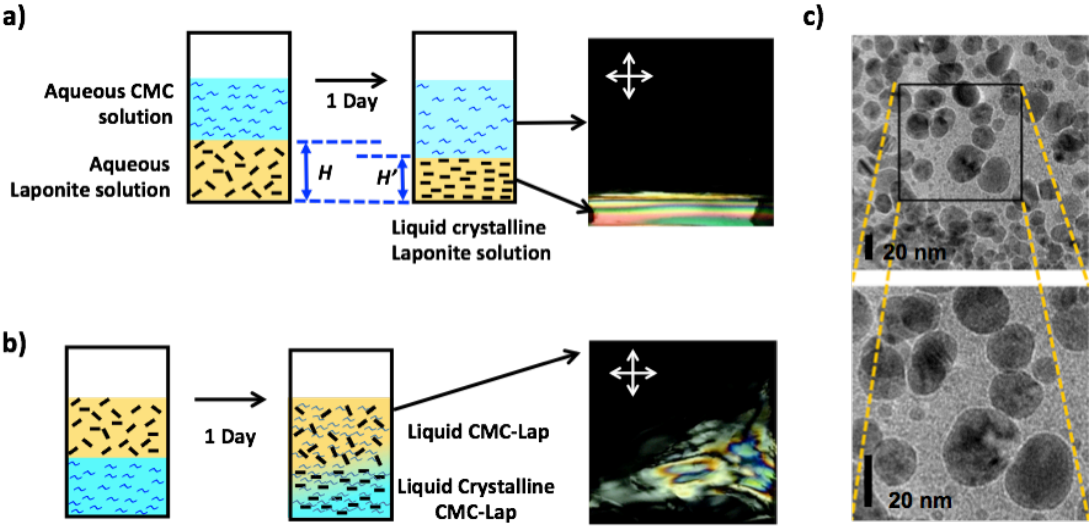

**Figure 2:** Schematic representation and photograph of the development of **(a)** a 0.8 wt% CMC solution placed carefully on top of a 2.0 wt% Laponite solution. After one day the interface between the two solutions shifts from an initial height H to a lower one at H' coinciding with the appearance of a birefringent texture, which is shown in the photograph take with crossed polarizers (the whit arrows indicate their polarization orientation). **(b)** Using the same volumes and concentrations, but placing the solutions in reverse order. **(c)** A TEM image taken from the liquid supernatant containing initially only Laponite particles after the sample has aged one day.

**Flexible and Transparent Thin Film**

The behavior of the clay and CMC mixture completely changed when we adopted a third mixing approach, namely stirring the sample for 24 h after adding a given amount of clay suspension to a 0.8 wt% CMC solution, thus assuring that all Laponite particles were well dispersed within the cellulose solution. The resulting non-birefringent, clear and viscous fluids were pored into petri dishes and left to dry at room temperature (Figure 3a). The films obtained were transparent, freestanding and highly flexible (Figure 3b). A Scanning Electron Microscope (SEM) image taken from the cross-section of the film showed multilayered in-plane structure with strong interconnectivity between each layer (Figure 3c). As evaporation proceeded, the polymer-clay network collapsed like a shrinking sponge due to hydrodynamic interactions.[45] Since the drying process was slow, clay particles were given enough time to self-assemble into a three-dimensional *brick-and-mortar* structure similar to nacre with the reinforcing Laponite particles glued to each other by the soft but toughening CMC. However, as the size of one Laponite disk is too small to be seen by SEM, only layered structure formed by the polymer-clay network were observed

For the purpose of improving the thermal and mechanical properties, which will be discussed below, we also investigated the effect of the addition of metal ions to the composite films. CMC contains carboxylate ions, which can be chelated with divalent cations ($Ca^{2+}$ or $Mg^{2+}$) through ion coordination.[46] We used $CaCl_2$ for its ready availability and non-toxicity compared to other divalent metal ions ($Cd^{2+}$ or $Cu^{2+}$). Moreover, calcium is also the main constituent of nacre.[47] During the preparation, we first cut a small piece of a CMC-Lap film and submerged it in the calcium chloride solution for several hours (Figure 3). By doing this, calcium ions were able to infiltrate the nanocomposite films. After the $CaCl_2$ bath, the film was taken out and rinsed with de-ionized water several times and then dried at room temperature. With the help of energy dispersive spectroscopy (EDS), we found that the calcium ions, marked by the red regions in Figure 3c, had been successfully introduced into the hybrid films. It should be mentioned that pure CMC films immediately dissolved once immersed in calcium chloride solution, so we did not do any tests for calcium coordinated CMC films. Note that adding the higher-valent cations directly to the initial aqueous polymer-clay mixture lead to a quick gelation and letting the gels dry only lead to turbid, low-quality films.

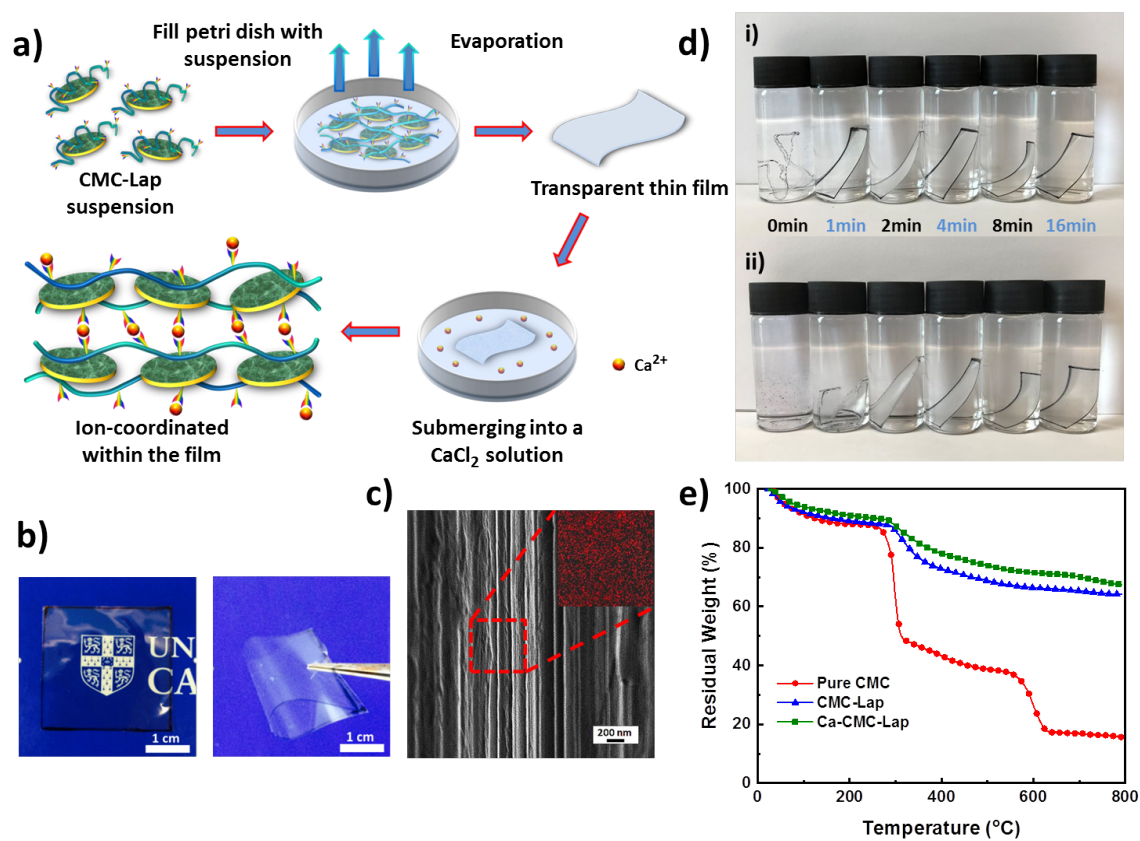

**Figure 3: (a)** A schematic representation of CMC-Lap hybrid film fabrication via evaporation and then subject to ion coordination through CMC further chelating with $Ca^{2+}$ ions. **(b)** Photograph showing the flexibility and transparency of the pure CMC-Lap film. **(c)** SEM image of the cross-sectional morphology of hybrid film showing well-ordered hierarchical structure with evenly distributed $Ca^{2+}$ ions as indicated by the red regions in the EDS image (inset). **(d)** Water resistance behavior of the pure and $Ca^{2+}$ coordinated CMC-Lap film obtained for varying dipping times in calcium chloride solution **(i)**. The black rim marks the transparent films. The pure CMC-Lap film swells in water while Ca-CMC-Lap films maintain their shape. Image **(ii)** shows the stability of the differently treated films after 20 days. **(e)** Thermal degradation profiles measured by thermogravimetric analysis of a pure CMC film, CMC-Lap and Ca-CMC-Lap hybrid film.

## Water- and Thermal-resistance Abilities

A simple experiment was carried out to compare the films' stability in water since CMC is a water-soluble polymer. For this, hybrid films with the same size and thickness were immersed in water (Figure 3d). Before submerging, they were subject to calcium ion coordination by placing them in a 120 g L$^{-1}$ CaCl$_2$ solution for different dipping times. Since our nanocomposite films were transparent, we marked their rims black to make them visible in water. Once the normal hybrid film (called CMC-Lap) touched the water, it immediately started to swell while those with ion coordination were still quite rigid initially. After 20 days, the film with only 1 min dipping time broke into small pieces while all the other films with longer dipping times maintained their original shapes. These experiments demonstrate that a roughly 5 µm thick samples needed only a few minutes of exposure to the CaCl$_2$ solution in order to reach their full strength and thus infiltration of the Ca$^{2+}$ ions into the entire film. This observation allows us to tune the water resistance of our nanocomposite films against disintegration, which may be of great use in the production of biodegradable materials. Note that the fully Ca$^{2+}$-coordinated CMC-Lap films to low and high pH solutions (ranging between pH 2 and 10) showed no deterioration either within 20 days.

To further investigate the thermal properties of our nano-composite films, we performed a thermogravimetric analysis (TGA). The results are shown in Figure 3e. Thermal decomposition of pure CMC generally occurs in two steps. The first step is related to weight loss, taking place between 100 and 250 °C, which is due to trapped water removal and the decomposition of the polymer side groups. In the second step between 250 and 600 °C, a further rapid weight loss is associated with the breakdown of the polymer backbone.[48,49] As for CMC-Lap films, the degradation rate decreases while residual weight increases, which is assigned to the confinement and thermal protection of the cellulose against oxidation by the clay particles. Interestingly, for CMC-Lap hybrid films with calcium ion modification, the degradation rate is further decreased. We ascribe the further increased thermal stability to the incorporation of metal ions into the nanocomposite film. Furthermore, compared to un-modified films, all those modified with calcium ion (Ca-CMC-Lap) retained a higher residual weight, which we attribute to the addition of the non-volatile calcium ions.

**Thin Film Characterization**

We further investigated the molecular interactions between the cellulose polymer and the clay particles and the addition of Ca$^{2+}$ ions by performing FTIR spectroscopy (Figure 4a). Neat Laponite displays two distinct peaks at 950 and 1099 cm$^{-1}$, which was attributed to the surface Si-O bond stretching.[50] The CMC spectrum displayed three characteristic peaks at 1322, 1417 and 1598 cm$^{-1}$, originating from C-O stretching vibration, symmetrical and asymmetrical C=O stretching vibrations of carboxylate ions respectively.[51] The broad and strong transmission band presented at 3403 cm$^{-1}$ was assigned to an -OH stretching vibration, corresponding to intermolecular and intramolecular hydrogen bonds within the CMC chains.[52] However, in the case of our polymer-clay hybrid film, the width of O-H band was broadened with weakened intensity, and the peak position red-shifted from 3403 to 3393 cm$^{-1}$ compared to pristine CMC. These results suggest weak hydrogen bonding between Laponite's silanol (Si-O) and CMC's hydroxyl (-OH) groups. In this way, the extent of hydrogen bonds formed between pure CMC was reduced. After calcium ion infiltration, peaks for asymmetrical and symmetrical carboxylate ions both shifted to lower values while the stretching band intensities were slightly weakened, indicating successful coordination of cellulose with calcium ions. Up to this stage, we had already combined the synergistic effect of both hydrogen bonds and ionic bonds.

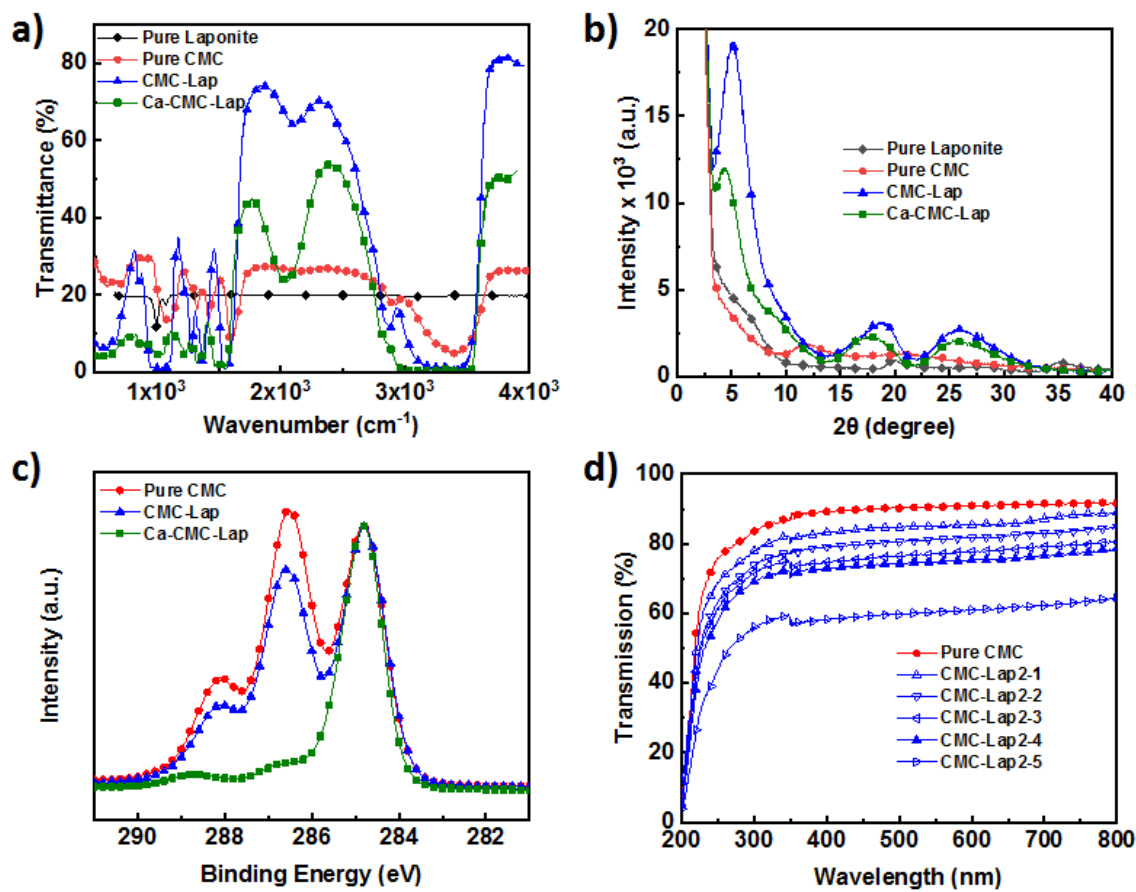

**Figure 4: (a)** FTIR and **(b)** XRD spectra of pure Laponite, pure CMC, CMC-Lap and Ca-CMC-Lap films. **(c)** C 1s core level XPS spectra of pure CMC, CMC-Lap and Ca-CMC-Lap and **(d)** UV-Vis transmission spectra taken from 10 μm thick CMC-Lap films with different CMC-Laponite weight fractions (as listed in the experimental section).

X-ray Diffraction was performed to determine the sub-nanometer structure of the composite films in order to see whether polymer intercalation into the spaces between clay particles was indeed successful. Using Bragg's law:

$$2d\sin\theta = n\lambda \quad (1)$$

where $\theta$ is the scattering angle, $\lambda$ is the wavelength of the incoming x-ray beam and $n$ is the order), characteristic interlayer distances, $d$, could be calculated from the peaks of the scattering intensity $I(\theta)$ shown in Figure 4b. In the case of neat clay powder, two broad and weak patterns were observed arising from the of small thickness of the clay platelets and their local crystallinity,[53] while pristine CMC films displayed two broad peaks in $I(\theta)$ with low intensity but different positions, which arise from its semi-crystalline morphology in its dry form.[54] For the CMC-Lap nanocomposite film, we observe enhanced peak intensities, which appear to be an overlay of scattering signals due to the pure Laponite and CMC at around $2\theta$ = 18° and 26°, while a much sharper, intense peak appeared at 5.06°, corresponding a characteristic length scale of 1.74 nm. We ascribe this peak to that fact that we indeed enveloped the 1nm thick Laponite particles with the cellulose chains. Upon drying these will rearrange into the proposed *brick-and-mortar* structure, explaining the apparently increased particle-to-particle distance, which is only weakly hinted by the weak but broad peak at around 6° in the pure Laponite scattering spectrum. It is interesting to observe that after the calcium ion intercalation, the strong first peak further shifted to $2\theta$ = 4.45° ($d$ = 1.98 nm), which agrees with our hypothesis that the $Ca^{2+}$ ions physically crosslink the cellulose chains trapped between the clay particles (see cartoon in Figure 3a). This also explains the additional broadening and shift of the two broad peaks at higher scattering angles seen for the Ca-CMC-Lap scattering curves in Figure 4b. However, there is also the possibility of the divalent calcium ions to directly bind between two clay platelets held together by the adsorbed polymer as they are negatively charged in water.

The physical binding between the polymer and the nanoparticle and the role of the $Ca^{2+}$ ions within the dried films was further studied with X-ray photoelectron spectroscopy (XPS). Typically, pure CMC will display 3 peaks at ~284.8 eV, 286 eV and 288.5 eV, which can be attributed to C-C, C–O, and C=O bonds respectively.[55,56] The C 1s core-level XPS spectrum (Figure 4c) of a composite CMC-Lap film reveals an apparent decrease in C-O signals compared to that of a pure CMC film, indicating the successful formation of the intermolecular hydrogen bonding between polymers and nanoparticles. The peak shift and a prominent signal decrease of C=O in the ion-coordinated film is due to the successful intercalation of calcium ions into the carboxyl groups in CMC biopolymers.

**Optical, Mechanical and Fire-retardent Properties**

Having established the local structure of our ion-coordinated nanocomposite films, we now present measurements testing the films' optical and mechanical properties as function of the clay-to-polymer ratios. First, we measured the transparency of the hybrid films using UV-vis spectroscopy (Figure 4d). The light transmittance of pure CMC was 85% at 650 nm. With the increase of clay to polymer ratio, the light transmittance slightly decreased especially for a CMC to Laponite ratio of 2:5 (CMC-Lap2-5). This decrease can easily be understood when considering the reflectance of the film in air:

$$R = |(1-n)/(1+n)|^2 \qquad (2)$$

where $n$ is the average refractive index of the composite film. The average refractive index of cellulose is 1.47 at 620 nm and that of Laponite is 1.54. Hence, as we increase the clay content the reflectance goes up, which is in accordance with the transmittance going down. Interestingly, the transparency of the nanocomposite films with different CMC-Lap compositions did not show significantly lower transmittance after the $Ca^{2+}$ intercalation, indicating that the ion coordination did not affect the films' transparency too much.

We also tested various mechanical properties of our nanocomposite films - the results are summarized in Table 1. Tensile stress measurements presented in Figure 5a show that the pure CMC film initially exhibited a linear elastic deformation before displaying a long plastic deformation region at low tensile stress. When the clay content was comparatively low (CMC-Lap2-1 to CMC-Lap2-2), Young's modulus, the maximum strength and toughness (strain-to-failure) were enhanced remarkably. For the nanocomposites with relatively higher clay content (CMC-Lap2-2 to CMC-Lap2-4), the mechanical properties improved further but at lower increasing rate. Indeed, the CMC-Lap2-4 films seemed to reach an optimal polymer-clay ratio as modulus and maximum tensile strength reached the highest values. Increasing the Laponite content even further finally lead to poorer mechanical properties (see curve for CMC-Lap2-5 in Figure 5). It is interesting to note that in contrast to the pure CMC films the hybrid films showed an increasingly shorter plastic region before sudden fracture, which occurred at around 4% applied tensile strain in the CMC-Lap2-4 sample.

**Table 1:** Thickness, Young's modulus, maximum strength and toughness of pure CMC, CMC-Lap hybrid and calcium-ion coordinated hybrid films

|  | Thickness (mm) | Young's Modulus (GPa) | Maximum Strength (MPa) | Toughness (MJm$^{-3}$) |
|---|---|---|---|---|
| CMC | 0.01 | 0.69 | 103.56 | 11.29 |
| CMC-Lap2-1 | 0.01 | 0.96 | 122.75 | 11.62 |
| CMC-Lap2-2 | 0.02 | 2.98 | 174.18 | 15.61 |
| CMC-Lap2-3 | 0.03 | 4.61 | 201.56 | 16.03 |
| CMC-Lap2-4 | 0.05 | 8.67 | 225.94 | 6.29 |
| CMC-Lap2-5 | 0.06 | 8.07 | 139.48 | 1.41 |
| Ca-CMC-Lap2-4 | 0.06 | 9.09 | 298.02 | 16.63 |

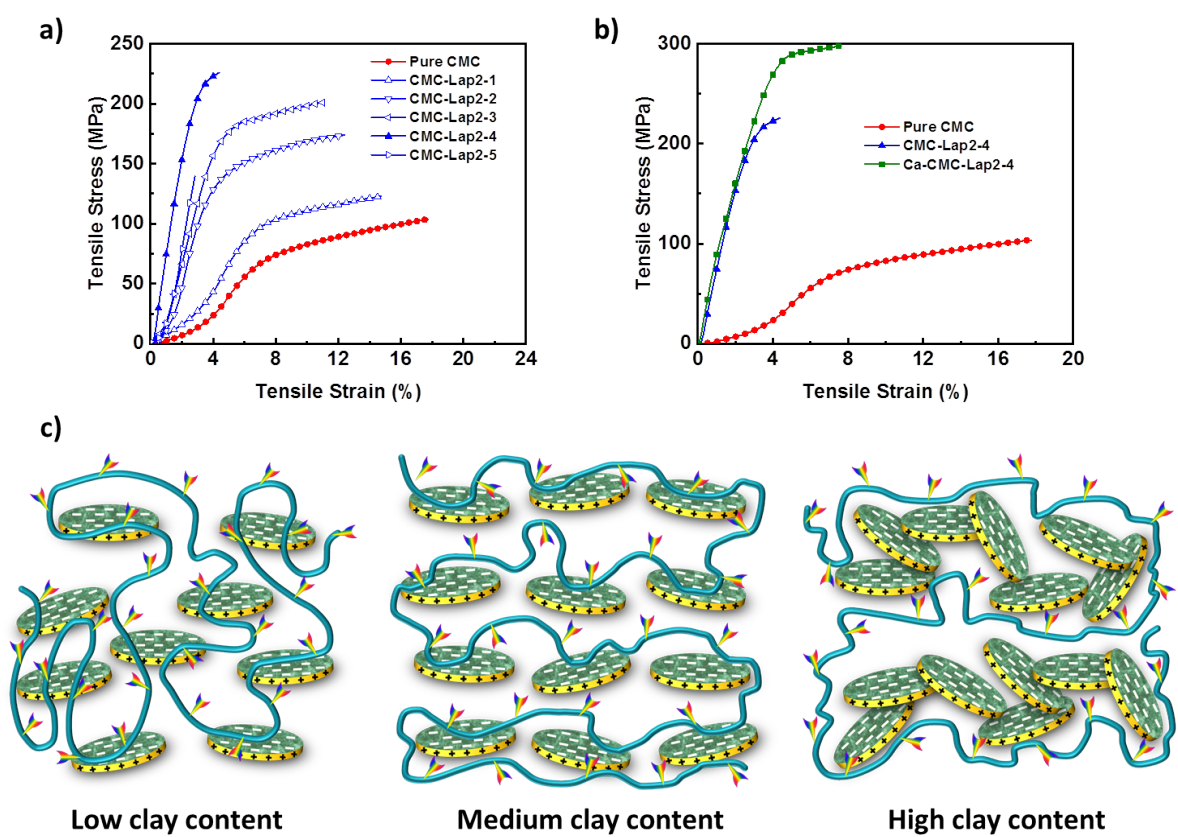

**Figure 5: (a)** Tensile stress-strain curves for pure CMC and CMC-Lap films with various polymer-clay weight ratios. **(b)** A comparison of mechanical properties between pure CMC, CMC-Lap2-4 and a calcium-ion modified CMC-Lap2-4 film. **(c)** Cartoon illustrating different clay contents.

To understand the change in the elastic and plastic regions of the pure polymer and the composite film one needs to remember that the pure CMC film is in fact a highly entangled polymer melt. Hence the strong plasticity over a large range of deformation (tensile strain) before the sudden rupture of the film occurs. The strength at lower deformations on the other hand comes from the intermolecular and intramolecular hydrogen bonding between segments of the cellulose chains.[57] However, when adding clay particles into the polymer matrix the degree of entanglement and therefore the range of plasticity will go down, but the adsorption to the clay particles and thus the increasingly strong bridging between them leads to reinforcing the strength of our hybrid films. Hence, compared to pure CMC films, more energy is required to break the hydrogen bonding at the polymer-clay interface and to induce interlayer slippage than for the disentanglement between the polymer chains.[58] Our measurements suggest that the polymer-to-clay ratio reached in the CMC-Lap2-4 samples delivers the maximum number of polymer to clay contacts in terms of perfect coating that delivers the best mechanical strength, similar to the *brick-and-mortar* structure found in nacre.[59] If using too much clay (CMC-Lap2-5), we run out of polymeric mortar to hold them together through adhesion and bridging, and if too few of the bricks are used the composite films will be dominated by the plastic behavior or polymer melts (CMC-Lap2-3 and lower ratios).

Having identified the CMC-Lap2-4 as the ideal mixture for making artificial nacre that exhibits an integrated high maximum strength (225.94 MPa), Young's modulus (8.67 GPa) and relatively high toughness (6.29 MJm$^{-3}$), we chose this composite films to study the effect of the calcium-ion coordination. Indeed, after dipping a CMC-Lap2-4 film into a $CaCl_2$ solution for 5 minutes and letting it dry, such modified films showed an even higher toughness (16.63 MJm$^{-3}$) and strength at break (298.02 MPa), which is 9.24 times and 2.29 times of nacre's toughness and maximum strength respectively (Figure 5b).[9,60] Surprisingly, even the plasticity range increased slightly. We suggest that this further improved mechanical behavior derives from the additional coordination bonding between the polymer and the calcium ion as well as the binding between two negatively charged platelets by the $Ca^{2+}$ ions that need to be broken when subject to external force. This suggests that ion coordination is able to further enhance the mechanical properties of our nanocomposite films while combining high strength and toughness at the same time.

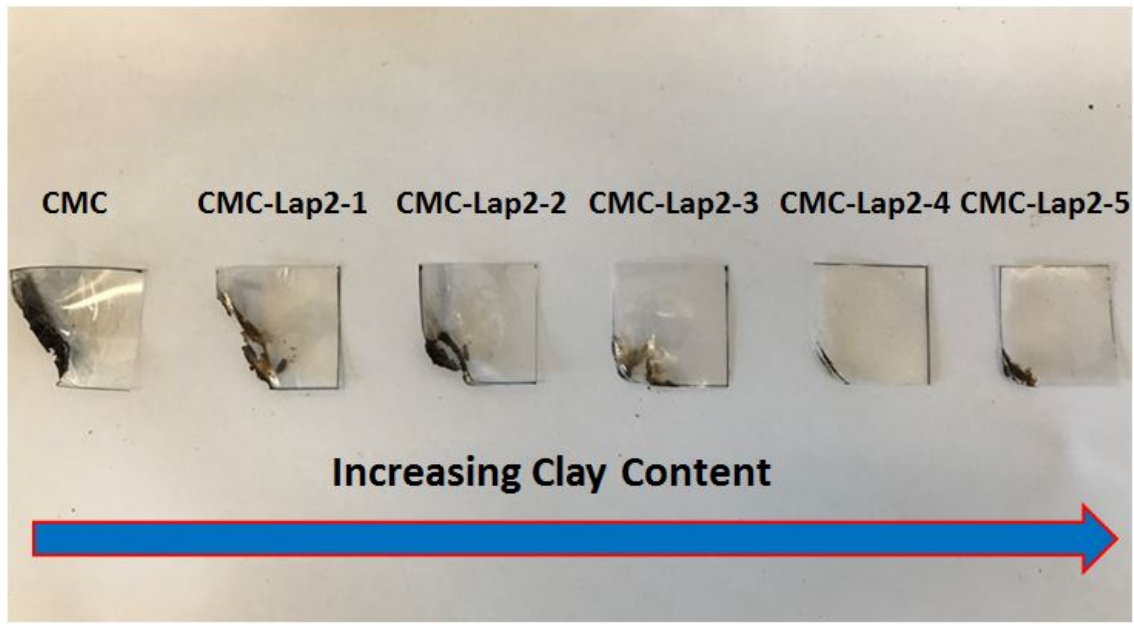

**Figure 6:** Photograph showing fire-retardant properties of PET films coated with pure CMC and CMC-Lap. With increasing added clay content the longer it takes to combust the PET film and its shape remained rather stable. In all cases we exposed the lower left corner of the coated films to a flame of a Bunsen burner.

Finally we tested the CMC-Lap composite films as possible fire-retardent coating of PET (polyethylene terephthalate) films, since PET is a very combustible resin that is widely used in people's daily lives.[61] A very thin layer of hybrid film was coated onto both sides of the PET films. The fire resistance ability was then estimated by burning these coated PET films. Upon exposing a corner of the coated film for 5 seconds to a Bunsen-burner flame, PET films coated only with cellulose burned with vigorous flames and dripped badly (Figure 6). However, PET films with nanocomoposite coating demonstrated improved fire-retardant ability. During the exposure to a direct flame, the film burnt initially only for a very short time because of small amount of polymers attached on the clay particles. After this, the specimen gradually became black, which was owing to carbonization of CMC.[62] When finishing burning out the surface polymer, the remaining clay particles did not catch any fire and remained inert. It needs to be pointed out that, unlike plastic materials, no dripping of hot fluids happened during the burning process. With increasing clay contents in the nanocomposite film coating, the char mass also increased along with less burning time and better-maintained dimension. As the clay content further increased, a completely different phenomenon occurred. The flame self-extinguished right after its removal and the coated PET could hardly be ignited. The reason for this outstanding fire-retardant property can be assigned to the densely packed layers consisting of the many inorganic clay particles. In general clays are known to be highly heat resistant and are excellent heat insulators. Hence, our polymer-clay coating can not only efficiently increase the thermal energy dissipation but also suppress the transport of heat and oxygen, which can then effectively protect the fresh PET film inside the coating. To summarize, $Ca^{2+}$-doped cellulose-Laponite composite films show remarkable optical transparency, mechanical strength and heat resistance, which can be utilized in the development of thin coatings that are environmentally friendly in production and recycling. Intriguingly, testing these different optical and mechanical properties by exposing the CMC-Lap films to solutions with other-valent cations such as $Fe_2(SO4)_3$, $AlCl_3$, $MgCl_2$ and KCl, the transparency, smoothness and with it the mechanical properties became inferior. Testing also different bio-polymers (sodium alginate, xanthan, chitosan and other cellulose derivatives) as the mortar between our clay-bricks did not deliver such superior, tunable properties as the CMC-Lap system.

## Conclusion

We have demonstrated a facile and simple pathway to fabricate superior bio-mimetic nanocomposite films resembling the structure of nacre but on nano-scale. These films are composed of the natural materials carboxymethyl cellulose and Laponite nanodisks. Through hydrogen bonding, the much longer polymers can attach to the clay surfaces thus forming trains of adsorption points on opposing clay particles leading to bridging inside the films with well-aligned architecture combined with a glass-like transparency and high flexibility. FTIR, XRD and XPS analysis proved that cellulose has been successfully intercalated into Laponite particles. By choosing the appropriate polymer-to-clay ratio we were able to obtain a 'perfect three dimensional *brick-and mortar*' structure that lead to very high toughness while the films remained transparent and flexible. The Young's modulus, maximum strength, and toughness were further enhanced by intercalating $Ca^{2+}$ cations between the clay particles, reaching 9.09 GPa, 298.02 MPa, and 16.63 $MJm^{-3}$, respectively. These values are higher than other nacre-inspired materials with the same inorganic content. The outstanding mechanical properties were supplemented by fire-retardant properties when used as protective coatings. Furthermore, the $Ca^{2+}$ treated nano-composite films showed a tunable resistance to dissolution in acidic and basic solution. We believe that in the near future, the scalable and environmentally friendly strategy along with the integrated high performance make these nanocompoiste materials promising candidates in different fields like transportation, wearable electronic devices, artificial muscles, aerospace and food packaging industries.

# Experimental

## Materials

Laponite XLG (Lap) was kindly provided by Rock Additives. Sodium carboxymethyl cellulose (CMC) with a molecular weight of 700 kDa and degree of substitution of 0.75 was purchased from Sigma Aldrich. Calcium Chloride anhydrous ($CaCl_2$) was also obtained from Sigma Aldrich. All the materials were analytical grade and used as received without further modification. All solutions were prepared using Milli-Q water.

## Preparation of Laponite and cellulose dispersions

A clear 0.8 wt% cellulose solution was prepared by dispersing cellulose powders into water, followed by mild stirring for 12 h at 90 °C. The Laponite suspension with varying concentration (0.4, 0.8, 1.2, 1.6 and 2.0 wt%) was prepared by adding a specific amount of dry Laponite powders, which were stored in a desiccator, into deionized water.[39,40] It should be mentioned that Laponite suspensions with higher concentrations were not prepared because they quickly aggregated and became gels.[39] In order fully disperse the Laponite powders in water, the solutions were vigorously stirred for 24 hours, and then filtered using a Millipore Syringe filter (from Sigma Aldrich) with a pore size of 0.22 µm.

**Preparation of Laponite-CMC liquid crystalline gels**

Our liquid crystalline gels were obtained by combining clay and polymer suspensions in different sequential orders. In one case, 5 mL 2.0 wt% Laponite solution was put in a glass vial and then 5 mL 0.8 wt% CMC was added slowly above it. In the other case, 5 mL 0.8 wt% CMC dispersion was added above a 5 mL Laponite dispersion (2.0 wt%) at the bottom of a glass vial. For both cases, glass vials containing clay-polymer liquid crystalline gels were sealed firmly and left for several days.

**Preparation of Lap-CMC nanocomposite films**

Our nanocomposite films were prepared in two steps. First, the clay suspension (5 mL) with specific concentration was added gradually into each cellulose dispersion (0.8 wt%, 5 mL) to form hybrids with CMC-Laponite weight ratios ranging from 2-1 to 2-5, named as CMC-Lap2-1, CMC-Lap2-2, CMC-Lap2-3, CMC-Lap2-4, CMC-Lap2-5, respectively. Subsequently the samples were subject to magnetic stirring for 24 h followed by an ultrasonic bath for 15 min for the purpose of removing air bubbles formed during mixing. As a control, a pure cellulose sample was prepared by diluting 5 mL CMC (0.8 wt%) with 5 mL deionized water to reach a concentration of 0.4 wt%. In the second step, all samples were transferred into petri dishes and allowed to evaporate at room temperature for 48 h. When evaporation finished, films were carefully peeled off from petri dishes.

**Preparation of Ca-CMC-Lap nanocomposite films**

Obtained dry films were cut into small strips (15 x 30 mm) and submerged into 120 g $L^{-1}$ calcium chloride solution. When finishing this, these calcium-ion-modified nanocomposite films were taken out and washed five times with deionized water and further dried at ambient conditions for 12 h.

**Characterization**

The cross section morphology of nanocomposite films was first coated with a conducting layer by sputtering it with a thin Au layer (3nm thick) and then imaged with a Scanning Electron Microscope (SEM) using a ZEISS Gemini EM with a build in energy dispersive spectroscopy (EDS) mode at an accelerating voltage of 10 kV. Transmission Electron Microscopy (TEM) images of cellulose coated Laponite particles were gathered using a Tecnai F20 S/TEM at liquid nitrogen temperature with an accelerating voltage of 200 kV. The sub-nanometer structure of the nanocomposite films was measured by X-ray diffraction (XRD) using a PANalytical Empyrean Series 2 Diffractometer System equipped with Cu-K$\alpha$ radiation ($\lambda$=1.54 Å) at 20 KV and 5 mA. Fourier-transform infrared (FTIR) examinations were carried out using a NICOLET iS10 FTIR spectrometer in the range of 500 to 4000 cm$^{-1}$. X-ray photoelectron spectroscopy (XPS) was conducted using an AXIS SUPRA photoelectron spectrometer (Kratos Analytical Ltd, UK) with a monochromatic Al-K$\alpha$ (hv = 1486.6 eV) X-ray source for excitation. The light transmittance of nanocomposite films was measured with a Cary 300 UV-Vis-NIR spectrometer. Thermal gravimetric analysis (TGA) was performed on a PerkinElmer Pyris 1 instrument, under nitrogen atmosphere with a heating rate of 20 °C min$^{-1}$ from 100 to 800 °C. The mechanical properties of the nanocompostie films and the reference polymer films were measured using a universal mechanical tensile tester (Instron 5500R-6025) equipped with a 200 N load cell at ambient conditions. For the mechanical testing, the films were first cut with scissors into rectangular strips of length 30 mm and width 15 mm. The film thickness (10-50 µm) was measured with a caliper. Then the films were mounted using tension clamps. The distance between the clamps was set to be 5 mm and a nominal loading rate of 0.5 mm min$^{-1}$ was applied. Five specimens were tested for each type of film to ensure the reproducibility. The integral below the stress-strain curves was used to calculate the toughness while the Young's modulus (E-modulus) was determined from the slopes in the linear elastic region of the stress-strain curves.

## Acknowledgements

We thank the Winton Program for the Physics of Sustainability for financial support. TE acknowledges the Royal Society for support via a Newton International Fellowship. We thank Linjie Dai for technical support on TEM and Zewei Li for doing the XRD measurements. We thank Thomas O'Neill for valuable discussions on interpreting the FTIR results.